\documentclass[11pt]{article}
\usepackage{amssymb}
\usepackage{amsmath}
\usepackage{graphicx}
\usepackage{color}

\def\beq{\begin{eqnarray}}    
\def\eeq{\end{eqnarray}}      

\newcommand{\Omo}{\Omega_m^0}

\newcommand{\OLo}{\Omega_{\Lambda}^0}

\newcommand{\rmr}{\rho_m}

\newcommand{\rL}{\rho_{\CC}}

\newcommand{\CC}{\Lambda}

\newcommand{\LQCD}{\Lambda_{\rm QCD}}

\begin{document}



 \hyphenation{cos-mo-lo-gi-cal
sig-ni-fi-cant par-ti-cu-lar}




\begin{center}
{\LARGE \textbf{Quantum Haplodynamics, Dark Matter and Dark Energy}} \vskip 2mm

\vspace{0.8cm}

\textbf{\large Harald Fritzsch\,$^{a,c}$,  Joan Sol\`{a}\,$^{b,c}$}

\vspace{0.5cm}

$^{a}$ Physik-Department, Universit\"at M\"unchen, D-80333 Munich,
Germany

\vspace{0.3cm}

$^{b}$ High Energy Physics Group, Dept. ECM and Institut de Ci{\`e}ncies del Cosmos\\
Univ. de Barcelona, Av. Diagonal 647, E-08028 Barcelona, Catalonia,
Spain

\vspace{0.5cm}

$^{c}$ Institute for Advanced Study, Nanyang Technological
University, Singapore

\vspace{0.25cm}

E-mails: fritzsch@mppmu.mpg.de, sola@ecm.ub.edu \vskip2mm

\end{center}
\vskip 15mm

\begin{quotation}
\noindent {\large\it \underline{Abstract}}.\ \ In quantum haplodynamics (QHD) the weak
bosons, quarks and leptons are bound states of fundamental
constituents, denoted as haplons.  The confinement scale of the
associated gauge group $SU(2)_h$ is of the order of $\Lambda_h\simeq
0.3$ TeV. One scalar state has zero haplon number and is the
resonance observed at the LHC. In addition, there exist new bound
states of haplons with no counterpart in the SM, having a mass of
the order of $0.5$ TeV up to a few TeV. In particular, a neutral
scalar state with haplon number $4$ is stable and can provide the
dark matter in the universe.

The QHD, QCD and QED  couplings can unify at the Planck scale. If
this scale changes slowly with cosmic time, all of the fundamental
couplings, the masses of the nucleons and of the DM particles,
including the cosmological term (or vacuum energy density), will
evolve with time. This could explain the dark energy of the
universe.

\end{quotation}
\vskip 8mm


\newpage

\vskip 6mm

 \noindent \section{Introduction}
 \label{Introduction}



The standard model (SM) of strong and electroweak interactions may
not be the final theory of the universe. The dark matter (DM) and
dark energy (DE) are fundamental problems awaiting for an
explanation\,\cite{CCproblem}. In this Letter we consider the
possible impact of the chiral gauge theory QHD (quantum
haplodynamics\,\cite{Fritzsch2012}) on these problems. QHD is a
theory of bound states for all the SM particles and the dark matter
particles.

If the QCD and QHD couplings take definite values around the Planck
scale, and this scale is permitted to slowly change with time, the
QCD and QHD couplings should change also with time. Hints that the
electromagnetic fine structure constant $\alpha_{\rm em}$ might
change with the cosmic evolution are reported in the
literature\,\cite{MurphyWebbFlaumbaum2003}. If $\alpha_{\rm em}$
changes in time, we expect that all fundamental coupling constants
change in time, including the gravity
constant\,\cite{FritzschSola2012,JSP-CCReview2013,Petersburg}.
{In particular, let us note that the particle masses
could also change in time, such as the proton mass. Different
experiments have explored this possibility, see e.g.
\cite{Reinhold06} and the reviews \cite{FundamentalConstants}.}
Since the gravity constant $G_N$ determines the Planck mass
$M_P=G_N^{-1/2}$, one expects that $M_P$ depends also on time and
slowly evolves with the cosmic expansion.

The present framework obviously implies a link between gravity and
particle physics. We shall come back to the cosmological
implications after describing the essentials of QHD.

\section{QHD} \label{sect:timeCC}

In QHD all of the SM particles (except the photon and the gluons)
are bound states of the fundamental constituents called haplons,
$h$, and their antiparticles. A first model of this type was
introduced in 1981 (see
\cite{FritzschMandelbaum,Barbieri,FritzschSchildknecht}). Here we
extend it and assume that the QHD chiral gauge group is the unitary
left-right group $SU(2)_L\times SU(2)_R$, which we will denote
$SU(2)_h$ for short.  All species of haplons $h$ are $SU(2)$
doublets, hence each one has two internal states $h_i$ represented
by the $SU(2)_h$ quantum number $i=1,2\,$. Rotations among these
states are performed by the exchange of two sets of massless
$SU(2)_h$ gauge bosons $\left(X_{L,R}^r\right)_{\mu}\,(r=1,2,3)$ for
each chirality.

There are six haplon flavors, two of them are electrically charged
chiral spinors ($\chi=\alpha,\beta$) and four are charged scalars
$S$. One scalar ($\ell$) has electric charge (+1/2) and carries
leptonic flavor. The other three scalars have charge (-1/6) and
carry color: $c_k=R,G ,B$ (``red, green, blue''). In Table I we
indicate the relevant quantum numbers.

The complete QHD gauge group is $SU(3)_c\times SU(2)_h\times
U(1)_{\rm em}$, with the coupling constants $(g_s,g_h, e)$. There is
a gauge coupling $g_h=g_h^L,g_h^R$ for each chiral factor in
$SU(2)_h$. The QHD part of the interaction Lagrangian involving a
generic chiral haplon $\chi$ ($\chi_L$ or $\chi_R$) and a scalar
haplon $S$ reads
\begin{eqnarray}\label{LSU3h}
{\cal L}_{\rm int}&=&\bar{\chi}^i\,\gamma_{\mu}i {\cal
D}^{\mu}_{ij}\chi^j+ \left({ D}^{\mu}_{ij} S^j\right)^{*}\,\left({{
D}_{\mu}}^{ij} S_j\right)+...
\end{eqnarray}
where ${\cal
D}^{\mu}_{ij}=\delta_{ij}\partial^{\mu}+ig_h^{L,R}\,\left(X_{L,R}^r\right)^{\mu}\left(\sigma_r/2\right)_{ij}$
is the $SU(2)_h$ covariant derivative and $\sigma_r$ are the Pauli
matrices. The other covariant derivative is coincident with the
previous one if $S$ is any scalar haplon.

For colored scalar haplons $c_k$, however, we have additional terms
where the covariant derivative is
$D^{\mu}_{ij}=\delta_{ij}\partial^{\mu}+ig_s\,A^{\mu}_a\left(\lambda^a/2\right)_{ij}$
and involves the gluons $A^{\mu}_a$ and Gell-Mann matrices
$\lambda^a$. Only the scalar haplons $c_k$ interact with the gluons,
since these are the only colored constituents. Besides the fine
structure constant $\alpha_{\rm em}=e^2/4\pi$ there is the strong
coupling $\alpha_s=g_s^2/4\pi$ and its $SU(2)_h$ counterpart
$\alpha_h=g_h^2/4\pi$, which is also strong at energies
$\mu\lesssim\Lambda_h$, but asymptotically free well above it. Here
$\Lambda_h$ denotes generically any of the two confining scales
$\Lambda_h^{L}$ and $\Lambda_h^{R}$ associated to the chiral factors
of $SU(2)_h$.

From the various haplon flavors the bound states of QHD can be
constructed. Only for energies $\mu$ well above  $\Lambda_h$  these
states break down into the fundamental haplons.  The weak gauge
bosons are s-wave bound states of left-handed haplons $\alpha$ or
$\beta$ and their antiparticles: $W^+  = \bar{\beta} \alpha$, $W^-
=  \bar{\alpha} \beta$ and $W^3  =\left(\bar{\alpha} \alpha -
\bar{\beta} \beta \right)/\sqrt{2}$.

The neutral weak boson mixes with the photon (similar to the mixing
between the photon and the neutral $\rho$-meson. One obtains the
physical $Z$-boson with a mass slightly heavier than the $W$-boson:
\begin{equation}\label{mixing}
\frac{M_Z^2-M_W^2}{M_W^2}=\frac{m^2}{1-m^2}\,,
\end{equation}
where the mixing parameter $m=e\,F_W/M_W$ is related to the $W$
decay constant $F_W$\,\cite{Fritzsch2012}. The confinement scale
$\Lambda_h^L$ for $SU(2)_L$ defines  the Fermi scale $G_F^{-1/2}\sim
0.3$ TeV and the size of the weak gauge bosons of the SM.

Owing to the  $SU(2)_L\times SU(2)_R$ chiral structure of QHD,
besides the three observed weak bosons there are also vector bosons
that couple to the righthanded leptons and quarks. We assume that
the righthanded confining scale $\Lambda_h^R$ is higher so that
the masses of the new vector bosons lie well above 1 TeV. They might
be observed in the new experiments at the LHC. The observed scalar
resonance is a p-wave excitation of the $Z$.  Here we will not
discuss these aspects of QHD in detail\,\cite{Fritzsch2012}.

The leptons and quarks are themselves bound states. They are
composed of a chiral haplon ($\alpha$ or $\beta$) and a scalar
haplon: $\ell$ for leptons and $c_k$ for quarks. The electron and
its neutrino have the structure ${\nu} =({\alpha}\bar{\ell})$ and
$e^-=({\beta}\bar{\ell})$, which is consistent with the quantum
numbers of Table I. Similarly, the up and down quarks (with $c_k$
color) are given by: $u =({\alpha}\bar{c}_k)$ and $d
=({\beta}\bar{c}_k)$.

In QHD the first generation of leptons and quarks describe the
ground states of the fermion-scalar bound states; the second and
third generation must be dynamical excitations. Likewise the c-quark
and t-quark families are excitations of the first quark generation.
Compared to the QHD mass scale the masses of the observed leptons
and quarks are essentially zero.

\begin{table}[t]
\begin{center}
\begin{tabular} {|c|c|c|c|c|}
  \hline
  $ \phantom{x}$ & $s$ & $Q$  & $SU(3)_c$ & $SU(2)_h$ \\ \hline\hline
  $\alpha$\phantom{x} & $1/2$ & $+1/2$ & $1$ & $2$ \\
  $\beta$\phantom{x} & $1/2$ & $-1/2$ & $1$ & $2$ \\
  $\ell$\phantom{x} & $0$ & $+1/2$ & $1$ & $2$ \\
  $c_k$\phantom{x}  &  $0$ & $-1/6$ & $3$ & $2$ \\
  \hline
\end{tabular}
\caption[]{Quantum numbers of the six haplons: spin ($s$), electric
charge $Q$ (in units of $|e|$) and corresponding representations of
$SU(3)_c$ and $SU(2)_h$.}
\end{center}
\end{table}


The outcome is an effective theory equivalent to the electroweak SM
in good approximation. However, new matter content is predicted. In
particular, the simplest neutral bound state of the four scalars
with haplon number ${\cal H}=4$ is a stable color singlet spinless
boson: $D =(lRGB)$. It is stable due to haplon number conservation,
which is similar to the conservation of baryon number.

\section{A New Dark Matter Candidate}\label{sect:DM Candidate}

The mass of the $D$ boson is expected to be in the region of a few
TeV. It can be produced together with its antiparticle by the
LHC-accelerator, and it can be observed by the large missing energy.
We interpret it as the particle providing the DM in the universe.
The properties of this DM particle are similar to a ``Weakly
Interacting Massive Particle'' (WIMP), but it can be much more
elusive concerning the interactions with nuclei.

After the Big Bang the universe is not only filled with a gas of
quarks and antiquarks, but also with a relativistic gas of
$D$-bosons and the corresponding antiparticles, which annihilate
into other particles. Due to the CP-violation there is an asymmetry
in the number of D-bosons and anti-D-bosons. After freezout, at
temperatures roughly $\sim 1/20$ of the $D$-boson mass, the $D$
particles abandon the equilibrium. A relic density remains today, a
gas of $D$ bosons, forming the dark matter. If we use the average
density of matter in our galaxy, we find that there should be ${\cal
O}(100)$ $D$-particles per cubic meter.

In contrast to more conventional WIMP's, the QHD candidate for DM
can escape more easily the recent, highly restrictive, bounds
obtained from scattering of DM particles off
nuclei\,\cite{LUXbounds}. It is not difficult to estimate the cross
section for the D-boson off a nucleon ${\cal N}$ (of mass $m_{\cal
N}$). It should be of order

\begin{equation}\label{QHDcross-section}
\sigma_{\scriptsize D{\cal N}}\sim
f_D^2\,\frac{\alpha_h^2}{\Lambda_h^4}\,m_{\cal N}^2\sim
f_D^2\,\alpha_h^2\,G_F^2\,m_{\cal N}^2\,,
\end{equation}
 where $G_F^{-1/2}\sim\Lambda_h\sim 300$ GeV  according to our definition of Fermi's scale in QHD. Here $f_D$ is the dimensionless form factor of the $D$-meson, which describes the confinement of the haplons by the $SU(2)_h$ strong gauge force. All QHD bound states have a form factor, which is of order one only for gauge boson mediated interactions, which are described by the exchange of weak bosons ($M_W^2\lesssim\Lambda_h^2$). For a deeply bound state as $D$, however, we rather expect $f_D\sim{\Lambda_h^2}/{B_D^2}\ll 1$, where $B_D$ is the characteristic binding energy scale.

In QCD the proton mass is given by a few times the value of the
confining scale $\Lambda_{\rm QCD}$, with only a tiny contribution
from the quark masses. Similarly, in QHD the masses of the bound
states are proportional to  $\Lambda_h$, although here the
spectroscopy is richer and the proportionality factor is bigger for
the more deeply bound states. Whereas for a weak boson such factor
is of order one, in the case of the DM candidate $D =(lRGB)$ it can
be much larger. This way we can {explain the small
cross-section of the DM particles with ordinary matter. For
$B_D>{\cal O}(10)$ TeV the scattering cross-section of $D$-bosons
off nucleons, Eq.\,(\ref{QHDcross-section}), can be approximately
reduced to the level of $\sim 10^{-45}$ cm$^2$ and become roughly
compatible with the current bounds\,\cite{LUXbounds}. Let us  note
that for energy scales $\mu>10$ TeV, we have
$\alpha_h(\mu)={\cal O}(0.1)$ owing to the asymptotic freedom of the haplon interaction constant (cf.
next section). One cannot be more precise at this point because at
present we cannot work out the details of the form factor $f_D$, and
therefore Eq.\,(\ref{QHDcross-section}) is only indicative of the
kind of result that could be obtained. The correct order of
magnitude could be reached if the bound state associated to DM in
this model is very tightly bound and massive. For the exceptional
four-haplon state $D =(lRGB)$ this situation should be regarded, in
principle, as possible. We believe that as a possibility it is
worthwhile to take into account, and still more considering that the present bounds from direct searches put very stringent constraints on virtually every DM candidate, including the more familiar ones from more conventional extensions of the SM.}

\section{Unification at the Planck Scale}\label{sect:GUT at MP}

The QHD, QCD and QED  couplings might unify at the Planck scale. We
can check it at one-loop level, starting from their low-energy
values and using the renormalization group equations (RGE's) to
compute the running of these parameters. For $SU(N)$ groups ($N>1$)
one has \,\cite{ChengEichtenLi74}:
\begin{equation}\label{RGE}
\frac{d\alpha_i}{d\ln\mu}=
-\frac{1}{2\pi}\left(\frac{11}{3}\,N-\frac23\,n_f-\frac16\,n_s\right)\,\alpha_i^2\equiv
-\frac{1}{2\pi}\,b_N\,\alpha_i^2\,,
\end{equation}

Here we have $\alpha_i=\alpha_h,\alpha_s$ ($n_f$ and $n_s$ are the
number of fermion flavors and scalars). For the $U(1)$ coupling
$\alpha_{\rm em}$ we have a similar formula as (\ref{RGE}), but in
this case
\begin{equation}\label{bU1}
b_{1}=-N_{h}\left(\frac43\,\sum Q_f^2+\frac13\,\sum Q_s^2\right)\,.
\end{equation}
Here $N_h=2$ for $SU(2)_h$. The electric charges $Q$ are defined in
Table I. For energies below $\Lambda_h$  we have to replace $N_h$ in
$b_1$ with $N_c=3$ (or $1$) and use the electric charges of the
quarks (leptons) rather than those of the haplons.

For the fine structure constant $\alpha_{\rm em}$ we extrapolate its
value from low energies to the Planck scale $M_P\simeq 1.22\times
10^{19}$ GeV. At the mass of the $Z$-boson we have $\alpha_{\rm
em}^{-1}(M_Z) = 127.94 \pm 0.014$. From the mass scale of the
$Z$-boson, $\mu=M_Z$, until a scale well above $\Lambda_h$, say
$\mu\sim 2$ TeV, we use the RGE, taking into account the charges of
the three charged leptons and of the five quarks, not including the
top-quark:
$
\alpha_{\rm em}^{-1} (2\, {\rm TeV}) = 123.57\,. $
From $\sim 2$ TeV up to the Planck mass we take into account the
electric charges of the two spin-$1/2$ haplons  and of the four
scalar haplons (cf. Table I). The result is:
\begin{equation}\label{alpha}
\alpha_{\rm em}^{-1}(M_P) = 114.58\,,
\end{equation}
equivalently: 
$\alpha_{\rm em}(M_P)=0.008727$ (cf. Table II).

\begin{table}[t]
\begin{center}
\begin{tabular} {|c|c|c|c|}
  \hline
  $ \phantom{x}$ & $\mu_0$ & $\mu_1$ &  $M_P$\\ \hline
  $ \phantom{x}$ & $M_Z$ & $2$ TeV & $10^{19}$ GeV  \\ \hline
  $\alpha_{\rm em}$\phantom{x} & \phantom{x}$0.007816$\phantom{x} & \phantom{x}$0.008092$\phantom{x} & \phantom{x}$0.008727$\phantom{x} \\
  $\alpha_s$\phantom{x} & $0.1184$ & $0.08187$  & $0.01370$\\
  $\alpha_h$\phantom{x} & $-$ & $0.62$  & $0.030$  \\
  \hline

\end{tabular}
\caption[]{The QED, QCD and QHD fine structure constants
$\alpha_i=g_i^2/4\pi$ at the $Z$-pole scale $\mu_0=M_Z$, at an
intermediate high energy scale $\mu_1=2$ TeV (around the haplon
continuum threshold), and at the Planck energy $M_P\sim 1.2\times
10^{19}$ GeV, for the $SU(2)_h$ chiral gauge group of QHD.}
 \end{center}
\end{table}

A similar procedure can be followed to compute the QCD coupling
constant at various energies. The accurate measurement of this
constant at the $Z$ pole yields: $\alpha_s(M_Z) = 0.1184 \pm
0.0007$. At the Fermi scale $\Lambda_h\sim 0.3$ GeV we find
$\alpha_s(\Lambda_h) = 0.1010$. Well above $1$ TeV up to the Planck
scale the renormalization proceeds via haplon-pairs (as indicated in
Table II).

For the $SU(2)_h$ group, we focus here on the lefthanded sector and assume $\Lambda_h \simeq 0.3$ TeV. We find e.g. $\alpha_h(2{\rm TeV}) =0.62$, and  eventually   
at the Planck energy:  $\alpha_h(M_P) \simeq 0.030$.   
From Table II we see that the three couplings approach each other at
the Planck scale. The details of the unification will depend on the
particular GUT group and can be affected by Clebsch - Gordan
coefficients of ${\cal O}(1)$.

We note that $SU(3)\times SU(2)_L\times SU(2)_R\times U(1)$ is a
natural breakdown step for GUT groups such as e.g. $SO(10)$. In our
case we do not have {spontaneous symmetry breaking (SSB)}, the
breaking is always meant to be dynamical. The complete QHD group can
thus be naturally linked to the GUT framework {without generating
unconfined vacuum energy}.

If the three couplings come close at the Planck scale, interesting
consequences can be derived in connection to the time variation of
the fundamental constants, of which hints in the literature appear
quite
often\,\cite{MurphyWebbFlaumbaum2003,Reinhold06,FundamentalConstants}.
An exact unification is not essential - we only need that the three
couplings take fixed values at or around $M_P$.

%


\section{Time evolution of Fundamental ``Constants''}\label{sect: Fundamental Constants}

A cosmic time change of Newton's constant $G_N$ (and hence of $M_P$)
is conceivable in the same way as one admits a possible time change
of $\alpha_{\rm
em}$\,\cite{MurphyWebbFlaumbaum2003,FundamentalConstants}. If the
QED, QCD as well as the QHD coupling constants emerge at the Planck
epoch, their primeval values should be very close and not be
time-dependent. Since the Planck energy changes in time, there must
be a time evolution of the gauge couplings at lower energies, say
around the confining scale of the weak bosons, $\Lambda_h\sim 300$
GeV. At the same time the masses of all the particles (including of
course the baryons and the $D$-bosons) are forced to slowly evolve
with the cosmic expansion since their binding energies are functions
of the coupling strengths.

We can estimate the time change of $G_N$ in QHD. We use the
approximate time variation of $\alpha_{\rm em}$ suggested in a
typical measurement where the current value of the QED coupling is
compared with that of a quasar some $12$ billion years ago
\,\cite{MurphyWebbFlaumbaum2003}: $\Delta\alpha_{\rm em}/\alpha_{\rm
em}=(-0.54\pm 0.12)\times 10^{-5}$.

From the RGE's and setting $\mu=M_P$ we can obtain the time
variation (indicated by a dot) of the Planck scale. Since
$b_1=-14/9$ in this case, we find
\begin{equation}\label{variationMPQED}
\frac{\dot{M}_P}{M_P}=-\frac{\dot{\alpha}_{\rm em}(M_Z)}{\alpha_{\rm
em}(M_Z)}\,\left[\ln\frac{M_P}{M_Z}+\frac{9\pi}{7\alpha_{\rm
em}(M_P)}\right]\,.
\end{equation}
It follows: $\Delta M_P/M_P\simeq 0.0027$ or ${\Delta G}/{G}\simeq
-0.0054$.

The time variation of the non-Abelian gauge couplings $\alpha_i$
(i.e. $\alpha_s$ and $\alpha_h$) at an arbitrary scale $\mu$ below
$M_P$ is also determined:
\begin{equation}\label{variationMPNonAbelian}
\frac{\dot{\alpha}_i(\mu)}{\alpha_i(\mu)}=\frac{\dot{M}_P}{M_P}\,\left[-\ln\frac{M_P}{\mu}+\frac{2\pi}{b_N\,\alpha_i(M_P)}\right]^{-1}\,,
\end{equation}
where $b_N$ was defined in (\ref{RGE}).

Since $\dot{M}_P/M_P$ is fixed from (\ref{variationMPQED}), the
above equation enables us to compute the cosmic time variation of
the QCD and QHD couplings  within the last $12$ billion years at any
desired energy well above $\Lambda_h$, e.g. at $\mu_1=2$ TeV (cf.
Table II):
\begin{equation}\label{Delta alphai}
\frac{\Delta{\alpha}_s}{\alpha_s}\simeq 1.1\times 10^{-4}\,,\ \ \
\frac{\Delta{\alpha}_h}{\alpha_h}\simeq 6.3\,\times 10^{-4}\,.
\end{equation}
Using the definition of the corresponding confining scales
$\Lambda_i$\, (viz. $\Lambda_{\rm QCD},\Lambda_h)$ we can check from
the above formulas that their cosmic time
evolution\,\cite{FritzschSola2012} is renormalization group
invariant and is directly tied to the cosmic evolution of $M_P$
itself:
\begin{equation}\label{linkalphaLambda}
\frac{\dot{\Lambda}_i}{\Lambda_i}=\frac{\dot{\alpha}_i(\mu)}{\alpha_i(\mu)}\,\frac{2\pi}{b\,\alpha_i(\mu)}=\frac{\dot{M}_P}{M_P}\,.
\end{equation}
Numerically, $\Delta\Lambda_i/\Lambda_i\simeq 3\times 10^{-3}$ for
the indicated period.

\section{QHD and Dark Energy}\label{sect: QHD and DE}

The unification of the QCD and QHD couplings can have a nontrivial
significance for the combined framework of particle physics and
General Relativity (GR). It suggests a cosmic evolution of all the
masses in the universe, both  of the nuclei and of the DM particles.
This can be perfectly compatible with  GR.

In order to preserve the Bianchi identity that is satisfied by the
Einstein tensor of the gravitational field equations
($\nabla^{\mu}G_{\mu\nu}=0$), the time evolution of the masses can
be compensated for by the time variation of one or more fundamental
gravitational parameters, typically the gravitational constant
$G_N$, or the cosmological constant $\CC$, or
both\,\cite{FritzschSola2012,JSP-CCReview2013,Petersburg}.

With the help of the Friedmann-Lema\^\i tre-Robertson-Walker (FLRW)
metric one finds that the most general local conservation law
preserving the Bianchi identity, corresponding to an isotropic and
homogeneous dust matter fluid with density $\rho_m$, reads:

\begin{equation}\label{BianchiGeneral}
\frac{G_N'}{G_N}\,(\rmr+\rL)+{\rho}'_m+\rho'_{\CC}+\frac{3}{a}\rmr=0\,,
\end{equation}
where $a$ is the scale factor and the primes denote derivatives of
the various quantities with respect to it. $\rL$ is the vacuum
energy density.

From Eq.\,(\ref{BianchiGeneral}) we can understand how a theory of
bound states at low energies can lead to a general evolution of the
vacuum energy density $\rL$ and Newton's coupling $G_N$ in
combination with the particle masses.

{Let us note that although many the studies that motivated the
possibility of having variable fundamental constants of Nature were
mainly focused on possible time variation of the fine-structure
constant from QSO absorption
spectra\,\cite{MurphyWebbFlaumbaum2003}, subsequent investigations
admitted the possibility of a time variation of the
particle masses, e.g. the proton mass\,\,\cite{Reinhold06}}. The predicted time evolution of the particle masses in QHD can be
parameterized as $\rmr\sim a^{-3(1-\nu)}$\,\cite{FritzschSola2012},
where the presence of $|\nu|\ll 1$ denotes a very small departure
from the standard conservation law $\sim a^{-3}$. Such a departure
is not viewed here as a loss or an excess in the number of particles
in a comoving volume (beyond the normal dilution law), but rather as
a change in the value of their masses. {Models with anomalous matter conservation laws of the above type have been carefully confronted with the precise
cosmological data on distant supernovae, baryonic acoustic oscillations, structure formation and CMB anisotropies, and one finds the upper bound $|\nu|\lesssim{\cal O}(10^{-3})$\,\cite{BPS09,BasPolSol12}.}

The anomalous matter conservation law  $\rmr\sim
a^{-3(1-\nu)}$ implies via Eq.\,(\ref{BianchiGeneral}) that a dynamical
response will be generated from the parameters of the gravitational
sector $G_N$
and $\rL$. 
This is how the dark energy can emerge: it is related to the change of the vacuum
energy density triggered by the time evolution of all the masses in
the universe. {While we cannot predict its value, we suggest that it is time dependent, which should be regarded as natural for the vacuum energy density of an expanding universe.}

{The time variation of the proton mass (and in general of all masses) within the aforementioned parameterization is approximately given as follows:}
\begin{equation}\label{eq:deltadotrho2}
 \left|\frac{\dot{m}_p}{m_p}\right|\simeq 3|\nu|\,\,H\,.
\end{equation}
{Notice that the index $\nu$ need not be universal. Here for simplicity we mention only the case of the proton\,\cite{FritzschSola2012}. The corresponding change of the vacuum energy density reads}
\begin{equation}\label{eq:deltaLambda}
\left|\frac{\dot{\rho}_{\CC}}{\rL}\right|\simeq -3|\nu|\,\frac{\Omo}{\OLo}\,H\,,
\end{equation}
{where $\Omo, \OLo$ are the current cosmological parameters associated to matter and vacuum energy.
Similarly, if the gravitational constant can change with time, we expect\,\cite{FritzschSola2012}}
\begin{equation}\label{GdotoverGo2}
\left|\frac{\dot{G}}{G}\right|\lesssim |\nu|\,H\,.
\end{equation}
{Using the
current value of the Hubble parameter as a reference, $H_0=1.0227\,h\times
10^{-10}\,{\rm yr}^{-1}\,$, where $h\simeq 0.70$, and the mentioned limit $|\nu|
{\lesssim\cal O}(10^{-3})$ we find that the time variations of the above parameters are at most of order $\lesssim10^{-13}{\rm yr}^{-1}\,$. These changes are of course only indicative since $\nu$ could be smaller, but at that level they are approximately within the present bounds. Such bounds are obtained from many sources and usually with large errors\,\,\cite{FundamentalConstants}. Let us mention that a new generation of high precision lab experiments could reach the level $\lesssim10^{-14}$
yr$^{-1}$\,\cite{Haensch}. In that case they might be sensitive to the possible time variations, if they are there.}

{One cannot exclude a priori that various sorts of effects are
involved at the same time. In such case the results can be more difficult to interpret and may require the use of more than one observable. In general this situation might enforce some reinterpretation of the previous observations, in which more emphasis is given to the variation of
one constant with respect to another. For example, it is interesting to note that in the context of conventional GUT's one finds that if
the $\LQCD$ scale (and hence the proton mass) changes with time, such change would be significantly larger (by more than one order of magnitude) than the possible time variation of the fine structure constant -- see
Ref.\,\cite{FritzschCalmet,FritzschSola2012}. At the end of the day the most
important feature is that a positive observational effect
(irrespective of the various sources of time variation concurring in
it) would signal the new qualitative fact that the constants of
Nature may be varying.}

{Finally, we should also mention that the models (\ref{BianchiGeneral}) where
the variable vacuum energy and gravitational constant can be linked to the variation of the
particle masses are compatible with the primordial nucleosynthesis
bounds on the chemical species. The important constraint to be
preserved here is that the vacuum energy density remains
sufficiently small as compared to the radiation density at the time
of nucleosynthesis. At the same time, potential variations of the gravitational constant should also be moderate enough to avoid a significant change in the expansion rate at that time. Once more one can show that these situations are under control provided the limit $|\nu|\lesssim{\cal O}(10^{-3})$ is
fulfilled\,\cite{BasPolSol12,GSFS10}. Under these circumstances the abundances of primordial chemical elements remain essentially the same.}

\section{Conclusions}\label{sect: Conclusions}

The framework we have outlined here proposes a new formulation of
GUT's involving gravity ab initio (in contrast to more conventional
formulations). It also proposes a new candidate for the dark matter
that is not in conflict with the recent, highly restrictive, bounds
for the scattering of DM particles off nuclei. The neutral and
stable $D$-boson is the particle of the dark matter.

QHD is not based on the conventional SSB mechanism and it does not
lead to a large contribution to the cosmological term. The DE
appears here as the tiny (but observable) dynamical change of the
vacuum energy density of the expanding background, and hence is a
part of the generic response of GR to the cosmic time variation of
the masses of all the stable baryons and dark matter particles in
the universe.

These ideas can be tested by future astrophysical and laboratory
tests in quantum optics, which are expected to detect potential
proton mass variations of order $\lesssim10^{-14}$
yr$^{-1}$\,\cite{Haensch}, hence at the level of the expected
running of the cosmological parameters.

\vspace{0.5cm}

\noindent{\bf Acknowledgments} \vspace{0.2cm}

\noindent HF thanks the BKC excellence programme and the Dept. ECM,
Univ. of Barcelona, for hospitality and support. JS is supported in
part by MICINN, DEC and CPAN. We are both grateful to the Institute
for Advanced Study at the Nanyang Technological University in
Singapore for hospitality and support.

\newcommand{\JHEP}[3]{ {JHEP} {#1} (#2)  {#3}}
\newcommand{\NPB}[3]{{ Nucl. Phys. } {\bf B#1} (#2)  {#3}}
\newcommand{\NPPS}[3]{{ Nucl. Phys. Proc. Supp. } {\bf #1} (#2)  {#3}}
\newcommand{\PRD}[3]{{ Phys. Rev. } {\bf D#1} (#2)   {#3}}
\newcommand{\PLB}[3]{{ Phys. Lett. } {\bf B#1} (#2)  {#3}}
\newcommand{\EPJ}[3]{{ Eur. Phys. J } {\bf C#1} (#2)  {#3}}
\newcommand{\PR}[3]{{ Phys. Rep. } {\bf #1} (#2)  {#3}}
\newcommand{\RMP}[3]{{ Rev. Mod. Phys. } {\bf #1} (#2)  {#3}}
\newcommand{\IJMP}[3]{{ Int. J. of Mod. Phys. } {\bf #1} (#2)  {#3}}
\newcommand{\PRL}[3]{{ Phys. Rev. Lett. } {\bf #1} (#2) {#3}}
\newcommand{\ZFP}[3]{{ Zeitsch. f. Physik } {\bf C#1} (#2)  {#3}}
\newcommand{\MPLA}[3]{{ Mod. Phys. Lett. } {\bf A#1} (#2) {#3}}
\newcommand{\CQG}[3]{{ Class. Quant. Grav. } {\bf #1} (#2) {#3}}
\newcommand{\JCAP}[3]{{ JCAP} {\bf#1} (#2)  {#3}}
\newcommand{\APJ}[3]{{ Astrophys. J. } {\bf #1} (#2)  {#3}}
\newcommand{\AMJ}[3]{{ Astronom. J. } {\bf #1} (#2)  {#3}}
\newcommand{\APP}[3]{{ Astropart. Phys. } {\bf #1} (#2)  {#3}}
\newcommand{\AAP}[3]{{ Astron. Astrophys. } {\bf #1} (#2)  {#3}}
\newcommand{\MNRAS}[3]{{ Mon. Not. Roy. Astron. Soc.} {\bf #1} (#2)  {#3}}
\newcommand{\JPA}[3]{{ J. Phys. A: Math. Theor.} {\bf #1} (#2)  {#3}}
\newcommand{\ProgS}[3]{{ Prog. Theor. Phys. Supp.} {\bf #1} (#2)  {#3}}
\newcommand{\APJS}[3]{{ Astrophys. J. Supl.} {\bf #1} (#2)  {#3}}

\newcommand{\Prog}[3]{{ Prog. Theor. Phys.} {\bf #1}  (#2) {#3}}
\newcommand{\IJMPA}[3]{{ Int. J. of Mod. Phys. A} {\bf #1}  {(#2)} {#3}}
\newcommand{\IJMPD}[3]{{ Int. J. of Mod. Phys. D} {\bf #1}  {(#2)} {#3}}
\newcommand{\GRG}[3]{{ Gen. Rel. Grav.} {\bf #1}  {(#2)} {#3}}



\end{document}